# Financial Performance and Economic Implications of COFCO's Strategic Acquisition of Mengniu

Jessica Ji [1,2], David Yu[3]
Jessicaji@uchicago,edu
David.yu@nyu.edu

[1] Department of Economics, New York University, New York, United States of America
[2] Harris School of Public Policy, The University of Chicago, Chicago, United States of America
[3] Department of Finance, New York University Stern School of Business, New York, United State of America

## 1 Introduction

### 1.1 Research Background

In 2008, the U.S. subprime mortgage crisis triggered a global financial crisis, which led to the collapse of many large enterprises and financial institutions, economic depression, and a significant increase in unemployment, as well as a sharp increase in global mergers and acquisitions (M&A), with large enterprises with strong capital in China going abroad to actively participate in overseas M&A (Ning Cui, 2019). At the same time, the "melamine" milk powder scandal occurred in China's dairy industry, which brought great harm to the nation's body and mind, a crisis to the dairy industry, and a series of negative impacts to China's dairy industry (Liqi Liu, 2017). At this time, COFCO Corporation ("COFCO") had been planning to enter the dairy industry for a long time, and had been actively looking for dairy companies that it could acquire to realize its own development strategy of the whole industry chain and to improve its international competitiveness to match that of the big international food companies. However, the dairy industry chain is very long, and it would not be easy for COFCO, which has been working on the plant chain for many years, to start from scratch, from laying out milk sources, building plants, and creating brands (Cao, J. X. & Dong, R. Chao, 2010). "After the melamine incident, China issued a series of laws and regulations and other related policies on dairy product safety, dairy industry access conditions, industry development and industrial transformation, etc. It was clearly proposed that state-owned enterprises should eliminate backward production capacity through mergers and acquisitions and restructuring of traditional enterprises, enhance innovation capacity, promote supply-side reform, and improve quality and efficiency. This provides an excellent opportunity for COFCO to enter the dairy industry (Liu, Litao., 2015). As a powerful grain, oil and food enterprise with a state-owned enterprise background, COFCO should not only assume the responsibilities and obligations of a state-owned enterprise, but also continuously strengthen itself, improve the layout of the global industry chain, and benchmark itself against the national standards of large grain merchants to be invincible in the international market competition (Xiapiao Xue & Wenxing Li, 2019). In such a context, the state-owned enterprises represented by COFCO need to develop rapidly through mergers and acquisitions in order to better practice the national strategic goals, and also to better realize the functions of state-owned enterprises. COFCO's acquisition of Inner Mongolia Mengniu Dairy (Group) Company Limited ("Mengniu") is based on such a background, and the contribution of Mengniu to COFCO is expected.

## 1.2 Research significance

By introducing the M&A process of COFCO and Mengniu Dairy, this paper studies the motivation and key points of this M&A, analyzes the financial and non-financial contributions of Mengniu to COFCO, and provides inspiration and reference for other corporate M&A activities. Therefore, this paper has strong theoretical and practical significance for the study of M&A or related M&A.

(1) Theoretical significance Through the collected data, it can be seen that the literature on M&A in the dairy industry is limited, especially the research on M&A contribution of dairy companies is less, and a perfect research system has not yet been formed. This paper takes the M&A case of COFCO and Mengniu as the research object to explore the issue of M&A contribution, and intends to analyze the contribution of this M&A activity to COFCO from two aspects: financial contribution and non-financial contribution, which broadens the research ideas of the M&A contribution issue and has positive theoretical value for the research in the field of M&A, and helps to enrich the theoretical research on the M&A contribution of dairy companies to a certain extent.

(2) Practical significance Corporate M&A is a complex task that requires companies to combine theory and practice and to integrate various resources and advantages of companies to carry out M&A activities. This paper analyzes the contribution of the M&A of Mengniu to COFCO based on the M&A case of COFCO and Mengniu. Through the analysis and summary of this M&A event, it can provide some inspiration and reference to other enterprises in China, especially dairy companies, in terms of the timing of M&A activities, the selection of target companies, the ways and means of M&A and the expected objectives.

## 2 Literature review

## 2.1 Research on the motivation of M&A

(1) Synergy effect theory
There are many studies on synergy effect. Synergy effect simply means that the earnings of two enterprises after merger and acquisition are greater than the sum of the earnings of two enterprises individually before the merger. According to the synergy theory, the synergy effect consists of operational synergy, financial synergy and management synergy. These three synergies are described in detail below.

For operational synergies, Ning Cui (2019), after studying the impact of M&A activities on the acquirer, concluded that corporate mergers and acquisitions can reduce production costs thereby increasing the firm's revenue and enhancing corporate value. The study focuses on whether M&A can achieve resource sharing and thus economies of scale.Xiapiao Xue & Wenxing Li (2019) state that M&A activities by firms in the same industry can achieve synergistic effects. Since M&A can be divided into horizontal and vertical M&A, this scholar believes that vertical M&A can help firms save costs.

For the study of financial synergy effect, Wei Weng,Lulu Sheng,Jieru Su & Lingyan Niu (2019) believe that M&A is a good way to solve the problem when enterprises are facing the situation of capital shortage, and he also suggests that the current financing mechanism in China needs to be further improved, and some enterprises have the problem of difficulty in financing, and through M&A activities can also The cost of financing can be reduced through M&A activities.

(2) Market power theory

From the market perspective, M&A activities can reduce the number of competitors in the same industry, and the M&A parties can expand their market share and increase their market share in this way. Jinling Jiang (2019)] found that M&A activities are also influenced by the market environment after related research. Since the motivation of M&A is closely related to the social environment in which it takes place, based on the special national conditions of socialism in China, the motivation of M&A of Chinese enterprises has its own special characteristics. Yahong Feng & Yunyun Wang (2018) found through their research that M&A activities have a phased pattern, for example, the M&A activities around 1997 in China have similar points, which is related to the regulation of the Securities Regulatory Commission at that time is inseparable. The study of M&A activities at a specific stage found that the purpose of M&A activities of listed companies to non-listed companies differed, both positively and negatively.

**2.2 Research literature on EVA analysis method**

Andriuskevicius Karolis & Ciegis Remigijus (2017) conducted a study using factor analysis by selecting common indicators in the financial indicators method and EVA indicators method, and the results of the study showed that EVA indicators method has more explanatory power, thus recommending companies to introduce EVA indicators method. The conclusion reached through the EVA indicator method is that the mergers and acquisitions not only do not achieve the value addition of the company in the M&A activities, but may even have a negative impact on the long-term development of the company.Liqi Liu (2017) found that the beneficiaries of the M&A activities are not the shareholders and the shareholders' value is not increased after analyzing the M&A cases of Chinese companies through the selected sample. Cao, J. X. & Dong, R. Chao (2010) concluded that the EVA indicator method is more objective and explanatory than the financial indicator method. Especially for a newly established company, it is normal to be in a loss-making situation, and the financial indicator method cannot reflect the development potential of the company, but the EVA indicator method can provide a more comprehensive understanding of the business performance of the company because it takes into account a wider range of factors.

## 2.3 Conclusion of Literature Review

As M&A activities are conducted more and more frequently, a large number of experts and scholars have intensified their research on M&A-related theories. The motivation of M&A is influenced by many aspects, synergies being the most common reason, but also by the market environment and national policies, etc. Also, since each case is unique, the motivation cannot be generalized. The most common evaluation method for M&A performance is the traditional financial metrics method, but with the increasing understanding and knowledge of the EVA metrics method, this method is gradually coming into view and has obvious advantages over the traditional financial metrics method, which takes into account shareholder value more comprehensively. However, there is no unified conclusion on the merits or demerits of the M&A performance evaluation method.

## 3 Analysis of M&A performance based on EVA method

The EVA research method is based on the comparison of the size of the economic benefits of a company conducting a merger or acquisition with the minimum payoff of the expected market returns, and provides a better measurement basis for assessing COFCO's M&A behavior of Mengniu. In the process of analyzing financial indicators, the cost of equity capital and the cost of debt capital are taken into account, while the possibility of internal manipulation of profits by listed companies is excluded by calculating net operating profit after tax. This section calculates the economic value added of COFCO's M&A of Mengniu from 2016-2020 through financial indicators to measure the wealth accumulation of the company's M&A of Mengniu.

I. Total profit

The total profit and the growth rate of total profit of the company for each year from 2016 to 2020 are collated and calculated in Table 3-1.

Table 3-1 COFCO's total profit from 2016 to 2020 (Unit: RMB million)

| Project | 2016 | 2017 | 2018 | 2019 | 2020 |
|---|---|---|---|---|---|
| Total profit | 1025478 | 1211445 | 1523658 | 1825475 | 2154783 |
| Growth rate | 16.35% | 18.13% | 25.77% | 19.81% | 18.04% |

Data source: Company annual reports

Net operating profit after tax (NOPAT for short) refers to the daily operating profit earned by a company after deducting income tax from its operating profit excluding interest expenses. The basic formula is Net Operating Profit After Tax = (Total Profit - Non-Operating Income + Non-Operating Expenses + Finance Costs + Impairment Loss on Assets + Development Expenses) x (1 - Income Tax Rate) - Increase in Deferred Income Tax Assets + Increase in Deferred Income Tax Liabilities, which can be calculated based on the financial indicators in the company's annual report.

II. Total invested capital

(i) Interest expense

According to Table 3-2, the highest interest expense of COFCO in the past five years is RMB 370 million and the lowest is RMB 240 million. The growth rate of interest expense is sometimes high and sometimes low, so the volatility has no reference value, so the arithmetic average of interest expense in the past five years is selected as the estimation parameter, and the value is taken as RMB 295 million.

Table 3-2 COFCO's interest expense from 2016 to 2020 (Unit: RMB million)

| Item | 2016 | 2017 | 2018 | 2019 | 2020 |
| --- | --- | --- | --- | --- | --- |
| Interest expense | 25478 | 36548 | 23658 | 28654 | 33258 |
| Growth rate | 25.36% | 46.54% | -51.42% | 28.32% | 21.25% |

(ii) Depreciation and amortization and capital expenditures

COFCO's depreciation amortization and capital expenditures for the period 2016-2020 are shown in Table 3-3.

Table 3-3 COFCO's depreciation amortization and capital expenditure for the period 2016-2020 (in RMB million)

| Item | 2016 | 2017 | 2018 | 2019 | 2020 |
| --- | --- | --- | --- | --- | --- |
| Depreciation and amortization | 236541 | 256547 | 276541 | 302547 | 335214 |
| Growth rate | 11.25% | 8.46% | 7.79% | 9.40% | 10.79% |
| Capital expenditure | 302147 | 325478 | 306987 | 332574 | 341258 |
| Growth rate | 5.63% | 7.72% | -5.68% | 8.33% | 2.61% |

Data source: Company's annual report

After calculation, the arithmetic average of the five-year depreciation and amortization increase rate of COFCO is 9.54% and the arithmetic average of the five-year capital expenditure is 3.72%.

III. Weighted Average Cost of Capital

Weighted average cost of capital (WACC) is the total cost of capital calculated as a weighted average of the various types of long-term capital costs of a company, using the proportion of each type of capital cost to the total capital as weights. It can be used to determine the rate of return required for investment projects with average risk. The basic formula is weighted average cost of capital ratio = debt capital ratio x cost of debt capital ratio + equity capital ratio x cost of equity capital ratio. The specific calculations are shown in Table 3-4.

Table 3-4 Projected depreciation and amortization and capital expenditure of COFCO for 2021-2025 (Unit: RMB million)

| Item | 2016 | 2017 | 2018 | 2019 | 2020 |
| --- | --- | --- | --- | --- | --- |
| Depreciation and amortization | 367193.4 | 402223.6674 | 440595.8 | 482628.6 | 528671.4 |
| Capital expenditures | 353952.8 | 367119.8 | 380776.7 | 394941.6 | 409633.4 |
| Cash Flow | 1254369.2 | 1247245.6 | 1215873.5 | 1185247.1 | 1254721.6 |
| Discount rate | 1.1145 | 1.0254 | 1.1547 | 1.2147 | 1.3157 |
| Present Value | 1397994.473 | 1278926 | 1403969 | 1439720 | 1650837 |

IV. Calculation of EVA value

According to the formula EVA＝NOPAY-TC* WACC, its calculation results are shown in Table 3-5.

Table 3-5: COFCO's capital operation from 2016-2020 (Unit: RMB million)

| Item | 2016 | 2017 | 2018 | 2019 | 2020 |
|---|---|---|---|---|---|
| NOPAY | 1035574 | 1311555 | 1533654 | 1435575 | 3155743 |
| WACC | 16.35% | 14.13% | 35.77% | 13.41% | 14.05% |
| TC | 236541 | 256547 | 276541 | 302547 | 335214 |
| EVA | 2530261 | 2603283 | 2868016 | 5154818 | 5430500 |

# 4 Performance analysis of M & A based on financial data

## 4.1 Solvency analysis

1. Asset liability ratio

It can be seen from table 4-1 that COFCO's asset liability ratio index has been high for a long time in the five years from 2016 to 2020, and the highest proportion reached 68.01% in 2020, indicating that COFCO's asset liability level has certain risks, the debt proportion in the company's asset structure is high, and the company's solvency for long-term debt is restricted to a certain extent. This is also closely related to COFCO's merger and acquisition of Mengniu. In order to cope with the huge financing pressure caused by the merger and acquisition of Mengniu, COFCO's main shareholders have cashed out their shares for many times, which not only worsened COFCO's asset liability ratio, but also aroused doubts about the motivation of shareholders' reduction. In this regard, COFCO needs to adjust the asset liability ratio to a level more acceptable to investors and the market according to the financial requirements of asset liability ratio control and referring to the average level of the same industry, so as to better have a positive impact on M & a performance.

2. Equity ratio

As can be seen from table 4-1, COFCO's equity ratio has been more than 1.5 times for a long time, and even more than 1.8 times in 2016 and 2020, indicating that COFCO has a deep degree of debt operation. Compared with shareholders' shareholding, the company makes more in-depth use of debt leverage. From the perspective of long-term debt length capability assessment, COFCO has a high equity ratio and faces great pressure on long-term debt repayment. Moreover, the debt level is too high and the company's long-term debt holding cost is large, which also easily leads to COFCO facing higher difficulties in the next round of financing, and the proportion of shareholders' equity is compressed too low. In the financing process, the necessary financial leverage and guarantee elements are inevitably in a dilemma. In fact, COFCO kept breaking the news of equity pledge financing by the company's main shareholders in 2020, indicating that COFCO's property right ratio control is facing a certain crisis. In this regard, COFCO needs to adjust and control the relevant indicators related to property right ratio, focusing on finance and financing. The company should reduce the equity ratio of the company on the premise of considering the pre tax deduction of interest expenses, so as to reduce the loan interest of the company, so as to alleviate the severe pressure of the company's financing costs and interest expenses on the solvency of long-term debt.

3. Interest cover

As can be seen from table 4-1, COFCO's interest coverage ratio from 2016 to 2020 is always greater than 1, indicating that the company's operating profit can cover the interest expense in the debt holding cost. Among them, the interest cover ratio in 2016 and 2017 was close to five times, and then directly jumped to nearly 15 times in 2018, which means that COFCO's operating profit has ushered in a great turnaround in 2018, and the huge increase in operating profit has strengthened the company's ability to repay the interest on long-term debt.

According to the comprehensive solvency analysis results, COFCO's short-term solvency and long-term solvency can basically reach the passing line of solvency indicators from 2016 to 2020. However, in terms of specific indicators, including but not limited to current ratio, quick ratio, equity ratio and other indicators, COFCO's solvency is actually facing considerable pressure. Combined with COFCO's development track and statements over the years, it can be found that the company has greatly expanded its fixed assets and production scale in recent years. Under this background, the deep application of financial leverage is inevitable. The dairy industry, especially COFCO's merger and acquisition of Mengniu, requires huge capital investment, its leverage utilization is generally high, and its solvency needs to be measured in a relatively long cycle. In this regard, COFCO needs to continue to use financial leverage to deliver capital for the company's long-term planning on the one hand, and pay attention to the balance of capital structure on the other hand to effectively protect the rights and interests of investors and creditors. The protection of creditors' interests can have a positive feedback effect on M & a performance and help optimize the development of M & A performance.

Table 4-1 Statistics of Main Long-Term Solvency Indicators of COFCO in 2016-2020

| Indicator / Year | Asset liability ratio | Property right ratio | Interest cover |
|---|---|---|---|
| 2020 | 60.07％ | 158.36 | 11.54 |
| 2019 | 68.01％ | 188.29 | 14.40 |
| 2018 | 64.07％ | 164.21 | 14.58 |

| 2017 | 60.08％ | 150.52 | 4.46 |
| 2016 | 64.41％ | 180.18 | 4.94 |

## 4.2 Analysis of operation capacity

(1) Turnover rate of current assets

As can be seen from Table 4-2, COFCO's turnover rate of current assets showed a wavy trend from 2016 to 2020, with an increase or decrease trend every other year. Among them, the highest year of turnover rate is 2018, reaching 1.8688, and the lowest year is 2016, only 1.1882. For an old comprehensive dairy enterprise, its liquid asset turnover is mainly realized through inventory turnover. With the full production of COFCO's manufacturing line in 2017, after the company's inventory management digests the backlog orders in the early stage, the coordination between production, manufacturing and market sales will directly lead to the change of the index health of current asset turnover. Overall, COFCO's current asset turnover rate shows an upward trend. With the continuous development of the company's development activities in the dairy market, the company's inventory management mechanism will be improved day by day. It is believed that COFCO's current asset turnover rate can maintain a steady increase.

(2) Turnover rate of accounts receivable

According to the changes in the index value of accounts receivable turnover rate in Table 4-2, COFCO's accounts receivable turnover rate showed a shortening trend from 2016 to 2020, which means that the average number of days COFCO's accounts receivable were converted into cash decreased year by year. Through the analysis of the reasons, it can be seen that the main business income in 2020 increased significantly, reaching 2 billion yuan. At the same time, the net accounts receivable also showed an upward trend on the whole. It shows that in recent years, with the expansion of business volume, the amount of credit sales of the company has expanded, resulting in a decline in the turnover rate of accounts receivable.

(3) Inventory turnover

As can be seen from Table 4-2, COFCO's inventory turnover rate showed an obvious upward trend from 2016 to 2020. The company's inventory turnover rate developed steadily from 2016 to 2018. The reason behind this is that COFCO has encountered the dilemma of limited production capacity in recent years. With the company's merger and acquisition in 2018 and the putting into use of dairy production capacity, COFCO's production capacity continues to climb, resulting in the rapid growth of the company's inventory turnover rate in 2019 and 2020, indicating that the cycle between COFCO's dairy production and sales is in good condition, and the inventory management cost under high turnover rate can be further diluted.

(4) Turnover rate of fixed assets

As can be seen from Table 4-2, COFCO's fixed asset turnover rate showed an overall upward trend from 2016 to 2020. The decline of the company's fixed asset turnover rate in 2017 was related to the centralized investment activities of fixed assets such as new production equipment and new plants at that time. After the integration and digestion of M & a capacity, COFCO's fixed asset turnover has been significantly increased in the short term.

(5) Total asset turnover

As can be seen from Table 4-2, COFCO's total asset turnover also maintained an increasing trend from 2016 to 2020. Compared with the compound annual average growth rate of 112% of the growth rate of fixed assets, COFCO's total asset turnover maintained a compound annual average growth rate of 40%.

As can be seen from the indicators of comprehensive operating capacity, COFCO's operating capacity has maintained a high-speed growth trend in general from 2016 to 2020. During 2017-2018, due to the company's integration after the merger and acquisition of Mengniu, there was a large decline in operating capacity indicators such as current asset turnover, fixed asset turnover and total asset turnover. However, with the continuous completion of integration, COFCO's operating capacity indicators have fully recovered and continued to maintain a high-speed growth trend. Whether it is current assets or fixed assets, COFCO's asset liquidity, asset utilization efficiency, asset turnover rate and other indicators operate well, which reflects COFCO's strong operating capacity.

Table 4-2 COFCO 2016-2020 annual operating capacity index statistics

| Year Indicator | 2020 | 2019 | 2018 | 2017 | 2016 |
|---|---|---|---|---|---|
| Current Assets Turnover Ratio | 1.819 | 1.4246 | 1.8688 | 1.5121 | 1.1882 |
| Accounts Receivable Turnover Ratio | 5.6886 | 4.9608 | 6.088 | 8.1444 | 8.4188 |
| Inventory Turnover Ratio | 4.8018 | 3.0211 | 2.5288 | 2.1828 | 2.8226 |
| Fixed Assets Turnover Ratio | 8.0052 | 8.2028 | 6.4228 | 8.1641 | 4.2258 |
| Total Assets Turnover Ratio | 1.2246 | 1.0194 | 1.2029 | 1.0688 | 0.8242 |

**4.3 Profitability analysis**

(1) Return on assets

As can be seen from table 4-3, COFCO's return on assets showed an overall upward trend from 2016 to 2020. It is worth noting that COFCO repeatedly carried out major project financing in 2018 and 2020, during which the total assets increased significantly and the valuation almost doubled. Nevertheless, in the face of a larger base of total assets, COFCO's asset return maintained a compound annual growth rate of 31%, indicating that COFCO's asset operation efficiency is good and the company has strong profitability.

(2) Return on net assets

As can be seen from table 4-3, COFCO's return on net assets showed a high growth rate from 2016 to 2020, and the compound annual growth rate of return on net assets reached 53%. While COFCO has made large-scale fixed asset investment on the one hand and financial leverage on the other hand, the return on net assets can achieve such a growth rate, reflecting the strengthening of COFCO's profitability. COFCO's early marketing expenditure and R & D investment have received good positive feedback after operating for a period of time. The company has continuously strengthened its profitability, continuously transfused blood for early losses and fixed asset investment, and continued to consolidate the stable development of the basic market of the company's capital structure.

(3) Gross profit margin of sales

As can be seen from table 4-3, COFCO's sales gross profit margin remained stable as a whole from 2016 to 2020, and the median value of gross profit margin hovered around 12%, which is also a feature of the dairy industry, that is, there was little change in cost and price. COFCO's marketing cost has always been large. The company attaches great importance to the laying of offline online stores, advertising and the construction of sales service system. At the same time, it has also made a large amount of investment and expenditure. Therefore, on the one hand, the BOM cost of dairy products manufacturing on the supply side is declining as a whole, on the other hand, COFCO's sales cost expenditure has occupied the profit control brought by the decline of manufacturing cost, As a result, the proportion of COFCO's sales cost expenditure in sales revenue has been difficult to decline, that is, the sales gross profit margin has been maintained at a relatively low level compared with other companies in the same industry for a long time, which poses a certain pressure on the long-term growth of COFCO's profitability. In this regard, COFCO needs to refine the management of sales cost expenditure and pay attention to the control of sales cost when planning sales expenditure, so as to improve the company's sales gross profit margin.

(4) Earnings per share (EPS)

As can be seen from table 4-3, COFCO's earnings per share changed little from 2016 to 2020, and the company's EPS operation was relatively stable. Generally speaking, the stock investment of listed companies mainly depends on the price difference formed by the change of stock price. Earnings per share is only the "icing on the cake" profit, but earnings per share is an important reference index of the company's profitability. COFCO's earnings per share have not made great progress in the past five years, but the company's share price has achieved a significant increase compared with the issue price, indicating that the market and investors are more optimistic about COFCO's development potential and optimistic about its long-term profitability.

Based on the analysis results of several indicators of profitability, it can be found that COFCO's overall profitability has been rising and strengthened from 2016 to 2020, and the company's profitability has been continuously enhanced, which not only enables the initial shareholders to obtain a higher book return on their investment in the initial stage of entrepreneurship, but also lays a relatively solid foundation for COFCO's subsequent series of highly leveraged mergers and acquisitions of Mengniu, This also enables COFCO to continue to gain the trust of the market when it falls into an operating crisis in 2020, and successfully get through the cash flow crisis by introducing external funds. Among them, COFCO's steady profitability is an important guarantee.

Table 4-3 COFCO's profit index statistics from 2016 to 2020

| Year<br>Indicator | 2020 | 2019 | 2018 | 2017 | 2016 |
|---|---|---|---|---|---|
| Return on assets (%) | 11.2143 | 10.8184 | 8.4817 | 8.1778 | 7.3141 |
| Return on net assets (%) | 31.2142 | 27.3443 | 24.8448 | 24.8887 | 18.2204 |
| Gross profit margin of sales (%) | 13.7408 | 13.4343 | 11.2484 | 11.8014 | 12.4742 |
| Earnings per share | 0.17 | 0.08 | 0.11 | 0.11 | 0.10 |

## 5 Conclusion

Both the unified financial index and EVA index show that COFCO's acquisition of Mengniu is a successful M & A, realizing the synergy of 1 + 1 > 2.

Mengniu Dairy's dairy industry is facing the adverse social impact caused by the melamine incident, and its brand reputation has been questioned by the public. Mengniu, whose main business is dairy products, urgently needs market integration to resolve the crisis. According to the annual report disclosed by Mengniu Dairy in 2008, the loss in that year was as high as 948 million, exceeding its full year profit in 2007. It can be seen that Mengniu Dairy is under great financial pressure. And continuously affected by the melamine incident and the OMP incident in 2009, the share price of Mengniu Dairy fell sharply. COFCO saw this great opportunity to acquire Mengniu, an excellent enterprise in the dairy industry at a low price, and made a decisive move to acquire Mengniu. On the surface, it is a financial M & A and an act taken to obtain short-term benefits. However, after more in-depth research, we found that Mengniu Dairy industry is an enterprise with long-term investment value. At the same time, we found from the existing literature that Mengniu Dairy has strong advantages in terms of profitability and operation ability in China's dairy industry, and has the potential to carry out effective cooperation with other enterprises. In addition, after a full analysis of the strategic plan established by COFCO group, it can be found that this M & A activity belongs to strategic M & A activity, in order to increase the synergy and achieve the goal.

**References**

[1] Ning Cui. (2019). Research on M&A Performance of Listed Companies in Internet Industry Based on Engineering Management. IOP Conference Series: Materials Science and Engineering(5). doi:10.1088/1757-899X/688/5/055050.

[2] Xiapiao Xue & Wenxing Li. (2019).Management Ability, Property Rights and Corporate M&A Performance — Empirical Evidence Based on Engineering Companies. IOP Conference Series: Materials Science and Engineering (5). doi:10.1088/1757-899X/688/5/055018.

[3] Wei Weng,Lulu Sheng,Jieru Su & Lingyan Niu.(2019).Data Analysis of M&A Performance in Automobile Industry——Taking Geely Auto as an Example..(eds.)Proceedings of 2019 International Conference on Educational Reform,Management Science and Sociology(ERMSS 2019)(pp.498-501).Clausius Scientific Press,Canada.

[4] Jinling Jiang.(2019).An Empirical Study on M&A Performance: Evidence from Horizontal Mergers and Acquisitions in the United States. Open Journal of Business and Management(2). doi:10.4236/ojbm.2019.72066.

[5] Yahong Feng & Yunyun Wang.(2018).Research on M&A Performance of Environmental Protection Enterprises under "One Belt and One Road". IOP Conference Series: Earth and Environmental Science(6). doi:10.1088/1755-1315/186/6/012012.

[6] Shan Xin,Hong Xie & Xiafei Jiang.(2018).Capital Mobility, Corporate Governance and M&A Performance——An Empirical Research Based on A-share Listed Companies in China..(eds.)Proceedings of the 2nd International Conference on Education,Economics and Management Research(ICEEMR-2018)(Advances in Social Science,Education and Humanities Research(ASSEHR),VOL.182)(pp.778-784).Atlantis Press.

[7] Andriuskevicius Karolis & Ciegis Remigijus. (2017).Developments and challenges of measuring M&A performance on a corporate and macroeconomic levels. Oeconomia Copernicana(2). doi:10.24136/oc.v8i2.13.

[8] Liqi Liu.(2017).Research on M&A Performance of Chinese listed TMT Companies..(eds.)Proceedings of 1st International Conference on Education,Economics and Management Research(ICEEMR 2017)(Advances in Social Science,Education and Humanities Research Volume95)(pp.77-80).Atlantis Press.

[9] Cui Zhe & Xie Jia.(2016).The Research of M&A Performance of Private Listing Corporation Based on Different M&A Incentives. International Journal of Economics, Finance and Management Sciences(5). doi:10.11648/j.ijefm.20160405.12.

[10] Guo Shengnan. (2018). A study on the contribution of the merger and acquisition of Mengniu to COFCO (Master's thesis, Harbin Institute of Technology).

Liu, Litao. (2015). Case analysis of COFCO's financing of M&A of Mengniu Dairy (Master's thesis, Institute of Fiscal Science, Ministry of Finance).

Sun Xiao. (2012). A case study of COFCO's M&A strategy and its effects (Master's thesis,

Beijing Jiaotong University).

Zheng, Y.M. & Yue, L.X. (2011). Analysis of new trends in corporate strategic management and practice from the perspective of mergers and acquisitions: A case study of COFCO. Mall Modernization (02), 12-13.

Cao, J. X. & Dong, R. Chao. (2010). A study on strategic M&A management based on financial perspective: A case study of COFCO's M&A of Mengniu Dairy. Finance and Accounting Monthly (07), 19-20. doi:10.19641/j.cnki.42-1290/f.2010.07.012.

Zhang Mo (2009-07-07). Will COFCO Houpao's joint venture in Mengniu bring a wave of M&A in the dairy industry? Economic Reference News, 003.

Appendix A: Financials

## APPENDIX A-1: INCOME STATEMENT

| In Php mn | 2018A | 2019A | 2020A | 2021E | 2022E | 2023E | 2024E | 2025E |
|---|---|---|---|---|---|---|---|---|
| Total revenue | 523,964.00 | 514,405.00 | 500,343.00 | 530,363.60 | 556,881.80 | 579,157.00 | 596,531.70 | 608,462.40 |
| Cost of goods sold | 394,605.00 | 385,301.00 | 373,396.00 | 397,772.70 | 417,661.30 | 434,367.80 | 447,398.80 | 456,346.80 |
| Gross profit | 129,359.00 | 129,104.00 | 126,947.00 | 132,590.90 | 139,220.40 | 144,789.30 | 149,132.90 | 152,115.60 |
| Selling, general and administrative | 108,791.00 | 107,147.00 | 106,510.00 | 91,752.90 | 96,340.50 | 100,194.20 | 103,200.00 | 105,264.00 |
| Net interest expense | 2,729.00 | 2,469.00 | 2,418.00 | 2,197.00 | 2,172.10 | 2,130.80 | 2,083.70 | 2,020.50 |
| Basic | 2,850.00 | 2,929.00 | 2,995.00 | 3,543 | 3,444 | 3,345 | 3,246 | 3,147 |
| Diluted | 2,868.00 | 2,945.00 | 3,010.00 | 3,547 | 3,448 | 3,349 | 3,250 | 3,151 |

# APPENDIX A-2:
# BALANCE SHEET

| Consolidated Balance Sheets (in US$ millions) | Actuals | | | Estimates | | | | |
|---|---|---|---|---|---|---|---|---|
| On January 31 | 2018A | 2019A | 2020A | 2021E | 2022E | 2023E | 2024E | 2025E |
| **Assets** | | | | | | | | |
| Current assets: | | | | | | | | |
|   Cash and cash equivalents | | 7,295.00 | 6,550.00 | 6,928.30 | 8,537.40 | 10,234.20 | 12,310.00 | 15,285.00 |
|   Receivables, net | | 5,189.00 | 5,937.00 | 6,482.20 | 6,806.30 | 7,078.60 | 7,290.90 | 7,436.80 |
|   Inventories | | 37,437.00 | 40,714.00 | 45,743.90 | 48,031.10 | 49,952.30 | 51,450.90 | 52,479.90 |
|   Prepaid expenses and other | | 2,860.00 | 1,685.00 | 2,752.60 | 2,890.20 | 3,005.80 | 3,096.00 | 3,157.90 |
|   Other current assets (discontinued operations) | | 121 | 89 | 89 | 89 | 89 | 89 | 89 |
|   Total current assets | | 52,902.00 | 54,975.00 | 56,129.00 | 60,193.80 | 63,953.30 | 67,638.00 | 71,717.70 |
| Property, plant and equipment, net | | 107,878.00 | 112,324.00 | 119,575.30 | 126,953.90 | 134,305.80 | 141,463.10 | 148,248.20 |
| Goodwill | | 15,763.00 | 20,651.00 | 20,651.00 | 20,651.00 | 20,651.00 | 20,651.00 | 20,651.00 |
| Other assets and deferred charges | | 4,229.00 | 5,456.00 | 5,486.00 | 4,869.00 | 4,602.00 | 4,632.00 | 4,015.00 |
| Total assets | | 180,772.00 | 193,406.00 | 200,301.30 | 209,564.20 | 218,847.80 | 228,189.90 | 236,937.40 |
| **Liabilities** | | | | | | | | |
| Current liabilities: | | | | | | | | |
|   Short-term borrowings | | 1,131.00 | 4,047.00 | 4,047.00 | 4,047.00 | 4,047.00 | 4,047.00 | 4,047.00 |
|   Accounts payable | | 32,676.00 | 36,608.00 | 41,766.10 | 43,854.40 | 45,608.60 | 46,976.90 | 47,916.40 |
|   Accrued liabilities | | 19,701.00 | 18,154.00 | 21,918.70 | 23,014.70 | 23,935.30 | 24,653.30 | 25,146.40 |
| | Accrued income taxes | | 157 | 1,164.00 | 704.2 | 738.5 | 764.6 | 781.1 | 787.4 |
| | Long term debt due within one year | | 4,455.00 | 1,975.00 | 1,975.00 | 1,975.00 | 1,975.00 | 1,975.00 | 1,975.00 |
| | Obligations under capital leases due within one year | | 326 | 326 | 326 | 326 | 326 | 326 | 326 |
| | Current liabilities of discontinued operations | | 47 | 26 | 26 | 26 | 26 | 26 | 26 |
| | Total current liabilities | | 58,493.00 | 62,300.00 | 63,967.00 | 66,845.80 | 69,261.20 | 71,141.40 | 72,427.50 |
| Long-term debt | | | 42,547.00 | 43,270.00 | 43,270.00 | 43,270.00 | 43,270.00 | 43,270.00 | 43,270.00 |
| Long-term obligations under capital leases | | | 3,154.00 | 3,109.00 | 3,109.00 | 3,109.00 | 3,109.00 | 3,109.00 | 3,109.00 |
| Deferred income taxes and other | | | 6,482.00 | 7,862.00 | 8,592.30 | 9,627.70 | 10,465.90 | 11,122.50 | 11,604.90 |
| Redeemable noncontrolling interest | | | 425 | 404 | 404 | 404 | 404 | 404 | 404 |
| Total liabilities | | | 111,101.00 | 116,945.00 | 120,028.00 | 123,910.20 | 127,116.70 | 129,593.20 | 131,290.70 |
| Shareholders' equity | | | | | | | | | |
| | Common stock par value + additional paid-in-capital | | 3,929.00 | 4,034.00 | 4,034.00 | 4,034.00 | 4,034.00 | 4,034.00 | 4,034.00 |
| | Retained earnings | | 71,967.00 | 78,691.00 | 66,897.20 | 55,372.70 | 44,293.70 | 32,850.10 | 21,903.90 |
| | Accumulated other comprehensive income (loss) | | 726 | -1,410.00 | -1,439.00 | -1,615.00 | -1,895.00 | -1,924.00 | -2,100.00 |
| | Total shareholders' equity | | 76,622.00 | 81,315.00 | 69,492.20 | 57,791.70 | 46,432.70 | 34,960.10 | 23,837.90 |
| | | Noncontrolling interest | 3,425.00 | 4,446.00 | 4,446.00 | 4,446.00 | 4,446.00 | 4,446.00 | 4,446.00 |

| | Total liabilities & equity | | 191,148.00 | 202,706.00 | 200,301.30 | 209,564.20 | 218,847.80 | 228,189.90 | 236,937.40 |
|---|---|---|---|---|---|---|---|---|---|
| **SUPPLEMENTAL DATA:** | | | | | | | | | |